\documentclass[prl,twocolumn,a4paper,showpacs,superscriptaddress]{revtex4}

\usepackage{graphicx}
\usepackage{amsmath}

\begin{document}

\title{New proposal for numerical simulations of
$\theta$-vacuum like systems}

\date{February 18, 2002}
\author{V.~Azcoiti}
\affiliation{Departamento de F\'{\i}sica Te\'orica, Universidad de
Zaragoza,
Cl. Pedro Cerbuna 12, E-50009 Zaragoza (Spain)}
\author{G.~Di Carlo}
\affiliation{INFN, Laboratori Nazionali del Gran Sasso, 67010 Assergi, 
(L'Aquila) (Italy)}
\author{A.~Galante}
\affiliation{INFN, Laboratori Nazionali del Gran Sasso, 67010 Assergi, 
(L'Aquila) (Italy)}
\affiliation{Dipartimento di Fisica dell'Universit\`a di L'Aquila,
67100 L'Aquila (Italy)}
\author{V.~Laliena}
\affiliation{Departamento de F\'{\i}sica Te\'orica, Universidad de
Zaragoza,
Cl. Pedro Cerbuna 12, E-50009 Zaragoza (Spain)}

\begin{abstract}
We propose a new approach to perform numerical simulations of 
$\theta$-vacuum like systems, test it in two analytically solvable 
models, and apply it to $\textrm{CP}^3$. 
The main new ingredient in our approach is the method used to 
compute the probability distribution function of the topological charge 
at $\theta=0$. We do not get unphysical phase transitions 
(flattening behavior of the free energy density) and reproduce the exact 
analytical results for the order parameter in the whole $\theta$-range 
within a few percent.
\end{abstract}

\maketitle

Quantum Field Theories with a topological term in the action are a
subject of interest in high energy particle physics and in solid state
physics since a long time. In particle physics, these models 
describe particle interactions with a CP violating term. The extremely
small experimental bound for the CP violating effects in QCD (strong CP 
problem) is still waiting for a convincing theoretical 
explanation \cite{strongcp}. 
In solid state physics, chains of half-integer quantum spins with
antiferromagnetic interactions are related to the two-dimensional $O(3)$
nonlinear sigma model with topological term at $\theta=\pi$. It has been 
argued that this model presents a second order phase transition at 
$\theta=\pi$, keeping its ground state CP symmetric 
(Haldane conjecture) \cite{haldane}.

Non-perturbative studies of field theories 
with a $\theta$-vacuum term are enormously 
delayed because of the complex character of the euclidean action 
which forbids 
the application of all standard Monte-Carlo algorithms.
Besides this, lattice QCD lacks from a simple consistent definition of 
topological charge. 

The partition function ${\cal Z}_V(\theta)$ of any $\theta$-vacuum like 
model in a finite space-time lattice volume $V$ can be 
written, up to a normalization constant, 
as the discrete Fourier transform of the probability distribution 
function (p.d.f.) of the topological charge at $\theta = 0$:
\begin{equation}
{\cal Z}_V(\theta)\;=\;\sum_n\,p_V(n)\,{\mathrm e}^{{\mathrm i}\theta n}\, ,
\label{partition}
\end{equation}
where $p_V(n)$ is the probability of the topological sector $n$ and the sum 
runs over all integers $n$. In almost all practical 
cases the sum in (\ref{partition}) 
has a number of terms of order $V$ since the 
maximum value of the topological charge at finite volume is of this order.

Since efficient algorithms for numerical simulations of physical systems 
with complex actions are not yet available, the only {\it a priori} reliable 
numerical scheme to analyze the thermodynamics of $\theta$-vacuum like 
models goes through the determination of the p.d.f. 
of the topological charge, $p_V(n)$, and the evaluation of its 
Fourier transform (\ref{partition}). But this is a difficult task due 
to the following two technical reasons, which will be clarified later: 
\textit{i}) 
any numerical determination of $p_V(n)$ suffers 
from statistical fluctuations \cite{PRD}, 
and small errors in $p_V(n)$ can 
induce enormous relative errors in the determination of a quantity as 
${\cal Z}_V(\theta)$ which is an extremely small number of order $e^{-V}$,
\textit{ii}) 
even if we were able to evaluate $p_V(n)$ with infinite precision, the 
sum in (\ref{partition}) contains terms that differ by many orders of 
magnitude, running from 1 to $e^{-V}$ \cite{RANDOM}.

In few specific cases one can overcome the sign problem \cite{STURN}. 
However 
previous attempts by other groups to simulate $\theta$-vacuum like systems 
\cite{GENER,PRD}, 
were based on the numerical determination of the p.d.f. 
of the topological charge straightforwardly,
by standard simulations, or by 
more sophisticated methods based on the use of multi-binning and re-weighting 
techniques. In all these attempts, 
artificial phase transitions at a $\theta_c$ 
decreasing with the lattice volume were observed for the two-dimensional 
U(1) gauge theory
at strong coupling as well as $\textrm{CP}^N$ models. 
The origin of this artificial
behavior, which follows from a flattening behavior of the free energy for 
$\theta$-values larger than a certain $\theta_c$, 
was analyzed in \cite{PRD,JAPAN}. Both groups 
agreed that the observed behavior was produced by the small 
statistical errors in the determination of the p.d.f. 
of the topological charge, the effect of which became more and more 
relevant as the lattice volume was increased. In Ref.~\cite{JAPAN} it 
was also noticed that by smoothing the p.d.f. flattening disappears. 

The purpose of this paper is to introduce a new numerical approach to simulate 
$\theta$-vacuum like models. This approach is based on a new method to 
compute the p.d.f. of the topological charge 
and the use of a multi-precision algorithm in order to compute the 
sum in (\ref{partition}) with a precision as high as desired. 

For reasons which will become apparent in what follows, let us write the 
partition function (\ref{partition}) as a sum over the density of topological 
charge $x_n =n/V$ and 
set $p_V(n)=\exp[-V f_V(x_n)]$, where $f_V(x)$ is a smooth
interpolation of $-1/V \ln p_V(x_n)$:
\begin{equation}
{\cal Z}_V(\theta)\;=\;\sum_{x_n}\,e^{-V f_V(x_n)}\,
{\mathrm e}^{{\mathrm i}\theta V x_n} \, .
\label{pfdiscrete}
\end{equation}
Equation (\ref{pfdiscrete}) defines a $2\pi$ periodic function of
$\theta$. Since CP is a symmetry of the action at $\theta=0$ and 
$\theta=\pi$, 
$f_V(x)$ will be an even function. 
We will assume that CP is realized in the vacuum at $\theta = 0$ since 
otherwise the theory would be ill-defined at $\theta\ne 0$ \cite{NOS}. This 
implies that $\exp[-V f_V(x_n)]$ will approach a delta distribution 
centered at the origin in the infinite volume limit. In 
some exceptional cases, as QCD in the chiral limit, the function $f_V(x)$ 
is not defined since any topological sector with non vanishing charge has a 
vanishing probability. However, this is a trivial case in which the 
theory is independent of $\theta$.

Let us consider the partition function (\ref{pfdiscrete}) in the
complex 
$\theta$
plane, in particular on the imaginary axis $\theta= -i h$, and let $f(x)$ 
be the infinite volume limit of $f_V(x_n)$. All the 
terms entering equation (\ref{pfdiscrete}) are positive definite, and 
then in the infinite volume limit the free energy is given by the
saddle point. Assuming that $f(x)$ has first derivative for any $x$ except 
at most in isolated points, we can write the saddle point equation:
\begin{equation}
f'(x) = h
\label{saddle}
\end{equation}
which gives the external ''magnetic field'' h as a function of the density 
of topological charge $x$.

Our proposal to compute the function $f(x)$ is based in the following 
three steps: 

\begin{itemize}
\item[i.] 
To perform standard numerical simulations of our system at imaginary 
$\theta = -ih$ and to measure the mean value of the density of topological 
charge as a function of $h$ with high accuracy (tipically a fraction of 
percent). This is feasible since the system we have to 
simulate has a real action. Then, Eq.~(\ref{saddle}) is used to get a 
numerical evaluation of $f'(x)$.  
\item[ii.] 
To get $f(x)$ we have to integrate $f'(x)$. 
Between the possible 
ways to do this integral, we decided to fit $f'(x)$ 
by the ratio of two polynomials, whose order is chosen to obtain a
high quality fit, and then to perform analytically the integral 
of the fitting function. In this way we get a very precise determination of 
$f(x)$, which allows us to compute the p.d.f. 
in a range varying several thousands 
orders of magnitude. 
This is the main advantage of our approach when compared with other 
methods based on a direct computation of $p_V(n)$.
\item[iii.]
To use a multi-precision algorithm to 
compute the partition function (\ref{pfdiscrete}) using as input the function 
$f(x)$ previously determined.
\end{itemize}

The function $f(x)$ obtained in step ii) suffers from statistical and 
systematic errors, the last coming from the fact that the saddle point 
equation (\ref{saddle}) has finite volume corrections. An analysis of these 
errors for the models and sizes we have studied (see below), 
shows that systematic 
errors due to finite volume effects are smaller than the statistical ones 
in the whole relevant range of $x$. This is the reason why  
we will replace $f_{V}(x_n)$ in equation (\ref{pfdiscrete}) by its asymptotic 
value $f(x_n)$ in what follows. This substitution has no effect in 
the infinite volume limit at imaginary $\theta$ and we are assuming that the 
same holds for real $\theta$.

Before presenting our results for the various testing models we have analyzed, 
let us briefly discuss how errors in the determination of $f(x)$ can 
propagate to ${\cal Z}_V(\theta)$. This is an important 
point since, as pointed out in \cite{PRD,JAPAN},
the artificial phase transitions found in $U(1)$ and 
$\textrm{CP}^N$ 
were caused by the statistical errors in the determination of the p.d.f.

\begin{figure}[t!]
\centerline{\includegraphics*[width=2in,angle=90]{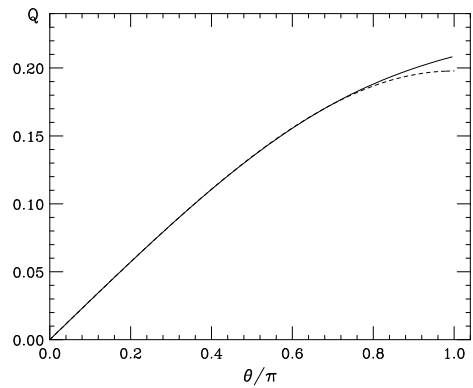}}
\caption{Magnetization \textit{versus} the imaginary magnetic field
in the one-dimensional antiferromagnetic Ising model at $F=-1/2$ on a
chain of 1000 sites: exact (dashed) and numerical (continuous) results. 
Statistical errors are smaller than $2\%$.}
\label{fig:ising_af}
\end{figure}  

To this end, let $f_V(x_n)$ be the exact value at a given lattice volume $V$ 
of the function which parameterizes the p.d.f. 
and be $\Delta f_V(x_n)$ a given 
deviation from the exact value. If we denote by ${\cal Z'}_V(\theta)$ the 
partition function computed with $f_V(x_n) + \Delta f_V(x_n)$, a simple 
calculation tells us that this partition function is related with the exact 
partition function ${\cal Z}_V(\theta)$ as follows:
\begin{equation}
{\cal Z'}_V(\theta)\;={\cal Z}_V(\theta) \langle e^{-V \Delta f_V(x_n)}
\rangle\, .
\label{error}
\end{equation}

We can at this point analyze two extreme cases. First let us assume that 
$\Delta f_V(x_n)$ vanishes everywhere except at a given $x_m$. Taking into 
account that the partition function ${\cal Z}_V(\theta)$ which enters 
in the denominator of the expectation value 
$\langle e^{-V \Delta f_V(x_n)}\rangle$ 
should behave as $e^{-V g_V(\theta)}$, where $g_V(\theta)$ is the free 
energy density ($g_V(0)=0$), we should get
\begin{eqnarray}
\langle e^{-V \Delta f_V(x_n)} \rangle \;=\; 1 \hspace*{5truecm} \nonumber \\ 
+\:2 e^{-V (f_V({m\over V}) - g_V(\theta))} 
(e^{-V \Delta f_V({m\over V})} - 1) \cos (m \theta)\, .
\label{error1}
\end{eqnarray}
Since $g_V(\theta)$ is an increasing function of $\theta$, it will be 
smaller than $f_V(x_m)$ near the origin, and therefore the free energy density 
computed with the modified partition function will differ from the exact 
one of a small quantity of order $1\over V$ in this region. 
However at larger values of 
$\theta$ the function $g_V(\theta)$ can become larger than 
$f_V(x_m)$ and in such a case the correction to the free energy density 
will be finite or, even worst, the modified partition function can become 
negative.
The other extreme case is that in which we assume that the error 
$\Delta f_V(x_n)$ is constant. Under such assumption Eq.~(\ref{error}) 
implies that the error in the free energy density $g_V(\theta)$ will also 
be $\Delta f_V$.

\begin{figure}[t!]
\centerline{\includegraphics*[width=2in,angle=90]{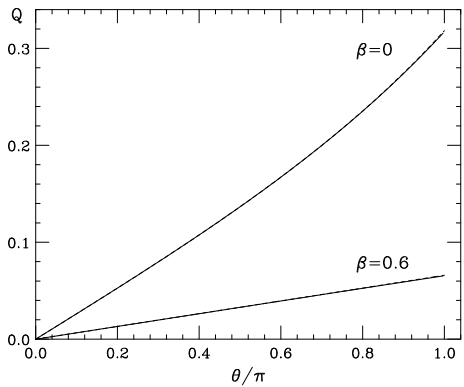}}
\caption{Topological charge \textit{versus} $\theta$ 
in the two-dimensional $U(1)$ model at $\beta=0$ and $\beta=0.6$
on a $80\times 80$ lattice. Statistical errors are not visible at this 
scale. 
The exact result (dashed curve) cannot be distinguished from the
numerical result (continuous curve).}
\label{fig:u1}
\end{figure}  

The previous discussion, as the results of \cite{JAPAN} 
suggests that correlated errors propagate 
in a less dramatic way than uncorrelated ones, and this would be 
a good scenario for numerical methods which, as in our approach, 
produce correlated errors in the determination of $f_V(x_n)$. We have 
checked this interesting issue in the two-dimensional U(1)
model at strong coupling (see below) 
and verified that in fact correlated errors in $f_V(x_n)$ induce errors 
of the same order in the free energy density $g_V(\theta)$.

To test these ideas we have analyzed three models:
the one-dimensional antiferromagnetic 
Ising model within an external imaginary magnetic field, the two-dimensional
compact U(1)  
model with topological charge, and $\textrm{CP}^3$ in two dimensions.
The coupling to the imaginary magnetic field in the Ising model
can be written as $\mathrm i \theta k_B T/2 \sum_i S_i$,
where $k_B$ is the Boltzmann constant and $T$ the temperature.
For an even number of spins, the quantity $1/2\,\sum_i S_i$ is an integer
taking all values between $-N/2$ and $N/2$, and therefore it can
be seen as a quantized charge. Furthermore, the 
theory has a $Z_2$ symmetry at $\theta=0$ and $\theta=\pi$ which 
is the analogous of CP in field theory. 
We use the notation $F=J/k_B T$,
where $J$ is the coupling constant between nearest neighbors. 
The transfer matrix technique allows to solve analytically the model.  
For antiferromagnetic couplings, $F<0$, the magnetization is an analytic 
function of $\theta$ between $-\pi$ and $\pi$. At $\theta = \pi$ the system 
shows a first order phase transition with a non-vanishing magnetization. 
From a numerical point of view the determination of the free energy density 
and order parameter through equation (\ref{partition}) in this model
has the same level of complexity of more sophisticated models. Furthermore, 
in contrast to 
two-dimensional $U(1)$ gauge theory, where the p.d.f. of 
the topological charge is nearly gaussian, the non-gaussian behavior of 
the p.d.f. of the mean magnetization in the 
antiferromagnetic Ising model makes this model a good 
laboratory to check the reliability of our
approach. Fig.~\ref{fig:ising_af} shows our 
numerical results for the order parameter versus $\theta$ for a linear chain 
of 1000 spins and $F=-1/2$. Statistical errors were estimated by doing 10 
samples of the numerical results and applying a Jack-Knife analysis.
The p.d.f. of the order 
parameter for such a system takes values in a range of around 2000 orders of 
magnitude. Notwithstanding that, we are able to reproduce the order parameter 
in the whole $\theta$ interval within a few percent.

\begin{figure}[t!]
\centerline{\includegraphics*[width=2in,angle=90]{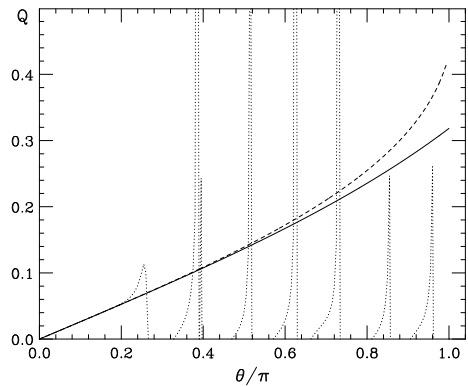}}
\caption{Topological charge \textit{versus} $\theta$ 
in the two-dimensional $U(1)$ model at $\beta=0$.
The continuous curve is the exact result, the dashed curve is the
result obtained by substituting $f(x)$ by
$f(x)(1+\frac{1}{2}\sin(x^2))$, and the dotted curve is the result
obtained by adding a random error of the order of 0.1\% to $f(x)$.}
\label{fig:errors}
\end{figure}  

The two-dimensional compact 
$U(1)$ gauge model with $\theta$-angle at strong coupling constitutes another 
interesting check
because we can compare the goodness of our approach with the other 
existing simulations which showed artificial 
behavior with a fictitious phase transition moving to the origin when 
increasing the lattice volume. 
Fig.~\ref{fig:u1} displays our results for the topological charge 
density versus $\theta$ 
in a 80$\times$80 lattice at $\beta=0$ and $\beta=0.6$. 
We are able to reproduce the exact result within a 
few per thousand in the whole $\theta$ interval. The agreement between 
analytical and numerical results is actually impressive. Furthermore the 
flattening found in \cite{PRD} for the free energy density in relatively 
small lattices is absent in our simulations even in the 80$\times$80 lattice.

To test how different kind of errors in the determination of the function 
$f(x)$ which defines the p.d.f. of the density of 
topological charge can affect the determination of the free energy and 
order parameter, we have added to the measured $f(x)$ a random relative 
error of order $10^{-3}$. Fig.~\ref{fig:errors} 
shows the order parameter obtained in this way. 
As can be seen 
a small but random error in $f(x)$ propagates to the order parameter in a very 
dramatic way and makes the calculation meaningless. Contrary to that if, 
in order to simulate a correlated relative error of order up to 50\%, 
we replace the measured $f(x)$ by the (even) function 
$f(x)\left(1+0.5\sin(x^2)\right)$, the result for the order parameter
is practically indistinguishable from the exact value for
$\theta<\pi/2$, and the maximum deviation is about 25\%, at $\theta=\pi$ 
(see Fig.~\ref{fig:errors}).
We conclude that random errors 
in $f(x)$ propagate in a very dramatic way but correlated errors do not, 
and this helps one to understand why our approach works so well.

The last model we have analyzed is $\textrm{CP}^3$ in two dimensional
euclidean space. It is the standard wisdom that this model shares many 
qualitative features with QCD. 
Even if it has not been analytically solved we 
believe it is worthwhile to compare our results with previous existing 
numerical simulations. 
We studied the lattice formulation that makes use of
an auxiliary U(1) field. 
Also in this model, the previous numerical simulations gave
artificial phase transitions 
with a flattening behavior for the free energy density at a 
$\theta_c$ decreasing with the lattice volume \cite{GENER,PRD}.

\begin{figure}[t!]
\centerline{\includegraphics*[width=2in,angle=90]{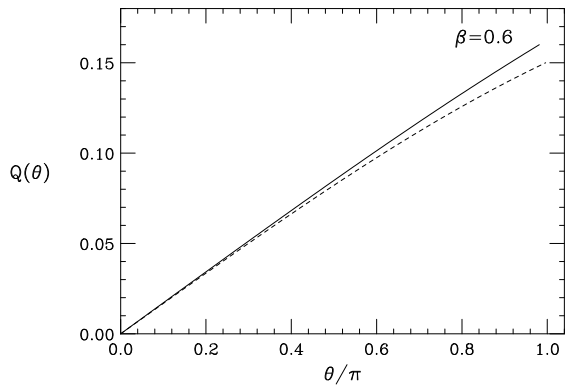}}
\caption{Topological charge \textit{versus} $\theta$ 
in the two-dimensional $\textrm{CP}^3$ model at $\beta=0.6$ on a
$100\times 100$ lattice. Continuous line (discontinuous line) reports 
the results obtained fitting $f'(x)$ with a polynomial (the ratio of two 
polynomials). Statistical errors are at $1\%$ level.}
\label{fig:cpn}
\end{figure}  

Fig.~\ref{fig:cpn} 
shows our results for the order parameter versus $\theta$ at 
$\beta=0.6$ on a $100^2$ lattice. 
We have chosen this particular $\beta$ value in order to compare 
directly with results reported in \cite{PRD}. In our simulations we have
no trace of the fictitious phase transition 
found in \cite{PRD}. Furthermore the order 
parameter is clearly different from zero at $\theta=\pi$, hence  
CP is spontaneously broken at this $\beta$ value. An open question is
how CP is realized in the continuum limit \cite{BURK}.
A more extensive analysis of this 
model, including a study of the behavior of the order parameter 
in the continuum limit, and an analysis of statistical and systematic errors 
involved in our approach, will be published elsewhere. 
What is interesting to notice here is that our 
results do not suffer from artificial phase transitions,
caused by statistical errors in the determination 
of the p.d.f. 

In the three models analyzed 
the finite size effects in $f'(x)$ cannot be appreciated 
since they are completely masked by the small statistical errors. 
For instance,
finite size effects can be exactly computed in the Ising model: they are 
exponentially small with the lattice size. This is a general feature
of non-critical systems. However, volume effects might be troublesome in
the analysis of the continuum limit.
Concerning systematic errors due to the choice of a particular 
fitting function for $f'(x)$, the difference between the numerical and exact 
results for the Ising and compact $U(1)$ models (beside the 
statistical errors) reported in Figs. 1 and 2 give us an idea on the order 
of magnitude of these errors. Of course systematic errors can depend on the 
model as well as on the parameters. In $CP^3$ at $\beta=0.6$ we did also 
a 5-parameters polynomial fit of the data for $f'(x)$ in the 
relevant $x$-interval.
The discrepancy between the topological charge density
obtained in the two cases is at most $7\%$ (see Fig. 4).

Similar ideas to those presented in this work have been proposed 
and promisingly applied to a matrix model of QCD at finite
density in \cite{NISH}.

This work has been partially supported by an INFN-CICyT collaboration and 
MCYT (Spain), grant FPA2000-1352. 
Victor Laliena has been supported by Ministerio de Ciencia y Tecnolog\'{\i}a 
(Spain) under the Ram\'on y Cajal program. Vicente Azcoiti and Victor Laliena 
would like to thank Professor Domingo Gonzalez for his invaluable 
contributions and very kind support.

\end{document}